\documentclass[aps,prd,floatfix,onecolumn]{revtex4} 

\usepackage[english]{babel}
\usepackage{amsmath}
\usepackage{amssymb}
\usepackage{graphicx}
\usepackage{dcolumn}
\usepackage{bm}
\usepackage{epstopdf}
\usepackage{slashed}

\usepackage{yfonts}

\usepackage{hyperref}

\usepackage{marvosym}
\usepackage{wasysym}

\newcommand{\myskip}[1]{}

\newcommand{\half}{{\frac{1}{2}}}

\newcommand{\tr}{{\rm tr}}

\newcommand{\BEQ}{\begin{eqnarray}}   
\newcommand{\EEQ}{\end{eqnarray}}   
\newcommand{\BEA}{\begin{eqnarray}}   
\newcommand{\EEA}{\end{eqnarray}}   
   
\renewcommand{\d}{{\rm d}}

\newcommand{\cD}{{\cal D}}

\newcommand{\cR}{{\cal R}}

\begin{document}


\begin{center}
{\Large \textbf{
Contra multos verbos: On scandals of quantum mechanics
}}
\end{center}

\begin{center}

Theodorus Maria Nieuwenhuizen
\end{center}

\begin{center}
{Institute for Theoretical Physics,   University of Amsterdam 
 \\ PO Box 94485, 1090 GL  Amsterdam, The Netherlands} \\   
* t.m.nieuwenhuizen@uva.nl
\end{center}



\section*{Abstract}{
In 2008 Nico van Kampen wrote in his letter  {\it The scandal of quantum mechanics}: 
``The scandal is that there are still many articles, discussions and textbooks,
which advertise various interpretations and philosophical profundities." 
Not much has changed since then, while social media have given a platform for more of what Nico would term ``a scandal''.

A detailed viewpoint is presented on the status of quantum mechanics, distilled from two decades of work with 
Armen Allahverdyan and Roger Balian on 
the dynamical solution of Curie-Weiss models for quantum measurement. It embodies a certain minimal form of the statistical interpretation and stays clear of ontological connections. Along the way, comments on various related subjects, terms and interpretations are given.

Contact to contributions of Andrei Khrennikov is made, to whom this essay is dedicated.
}




\tableofcontents

\section{Introduction}

\hfill{\it Congrats!} 

\hfill{{\it Half way-i,}

\hfill{\it Andrei \hspace{-1.8mm} !}
\vspace{3mm}

This essay is dedicated to the 60th birthday of Prof. Andrei Khrennikov.
We have shared some history, a warm friendship and deep mutual understanding. 
We met when Andrei invited me to the 2004 edition of his V\"axj\"o conference series on Foundations of Quantum Mechanics (and Probability),
and I have participated basically every time since then.
It was great to realize that we share our basic view on quantum mechanics (QM)
and it was great to find at the V\"axj\"o meetings a podium and a specialized audience for works on dynamics of quantum measurement 
and, from that, the interpretation of QM. The various questions led to further consideration and better formulations,
resulting in deeper understanding. Another effect, of course, is spreading the good news about steps of progress.

I would like to recall the 2005 V\"axj\"o meeting. I was co-organizer and celebrated my 50th birthday during the meeting.
On a sunny afternoon we made a cultural tour that ended by boat at an island. Andrei offered me a special gift, one of his books, 
and I toasted with the participants. Next day, the conference dinner in the splendid V\"axj\"o castle, was on my birthday.

 Andrei and I  coauthored a paper \cite{allahverdyan2005brownian} and we co-organized or participated in various meetings, 
notably the FQMT series in Prague, and Andrei lectured at summer schools I organized in Jo\~ao Pessoa.
So Andrei, thanks for that; I wish you a bright future and hope that the ideas I present here inspire you for further modelling, analysis and discussion.
Fig. 1 shows us at another V\"axj\"o meeting.

\begin{figure}\label{figLoic1}
\centerline{ \includegraphics[width=7cm]{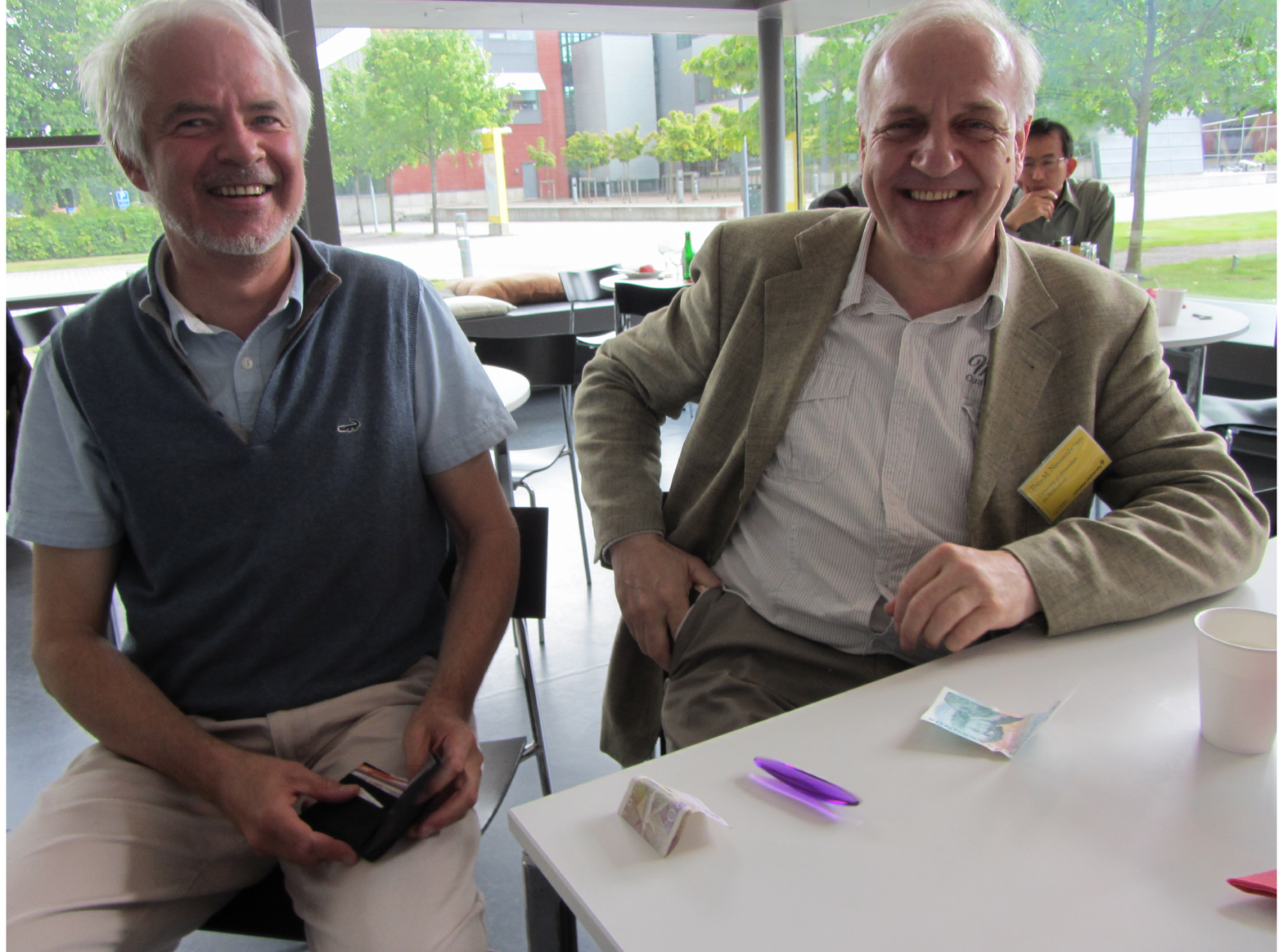}}
\caption{Andrei (left) and the Author during the 2012 V\"axj\"o conference, amazed by an asymmetric top.
If it is turned clockwise it continues its motion; if turned counterclockwise it stops and starts slowly turning clockwise. 
Courtesy by Marian Kupczynski.
}
\end{figure}

\newpage

\subsection{Oh, there is a scandal?}

\hfill{\it I read the news today, oh boy}

\hfill{\it And though the news was rather sad}

\hfill{\it Well, I just had to laugh \ldots}

\hfill{A Day in the Life}

\hfill{The Beatles}

\hspace{3mm}

The title of this essay is inspired by the 2008 letter to the editor of the American Journal of Physics with title
 {\it The scandal of quantum mechanics} by Nicolaas `Nico' Godfried van Kampen\cite{van2008scandal}.
Next to Martinus Veltman and Gerardus 't Hooft, Nico was one of my main teachers and influencers 
at the University of Utrecht in the 1970's. We coauthored a criticism \cite{nieuwenhuizen1987objections},
while our roads regularly crossed until his passing in  2013. 
Nico published on the interpretation of QM in 1958 and 1988.  
The latter paper, titled {\it Ten theorems about quantum mechanical measurements}\cite{van1988ten}, 
can be summarized as: Theorem I: {\it Quantum mechanics  works} and Theorem IV:
{\it Whoever endows $\psi$ with more meaning than  is needed for computing  observable phenomena 
is responsible for the consequences. }
I remember Nico presenting it at the annual meeting of the Netherlands Physical Society in the early 1990s, a vivid presentation in a fully packed room.
Nephew Gerard(us) came in late and had to sit on the floor, as he must have done so often as a boy, 
trying to grasp as much as possible from his uncle's lessons. For a photograph of Nico, see Fig. 2.

\begin{figure}\label{figLoic1}
\centerline{ \includegraphics[width=6cm]{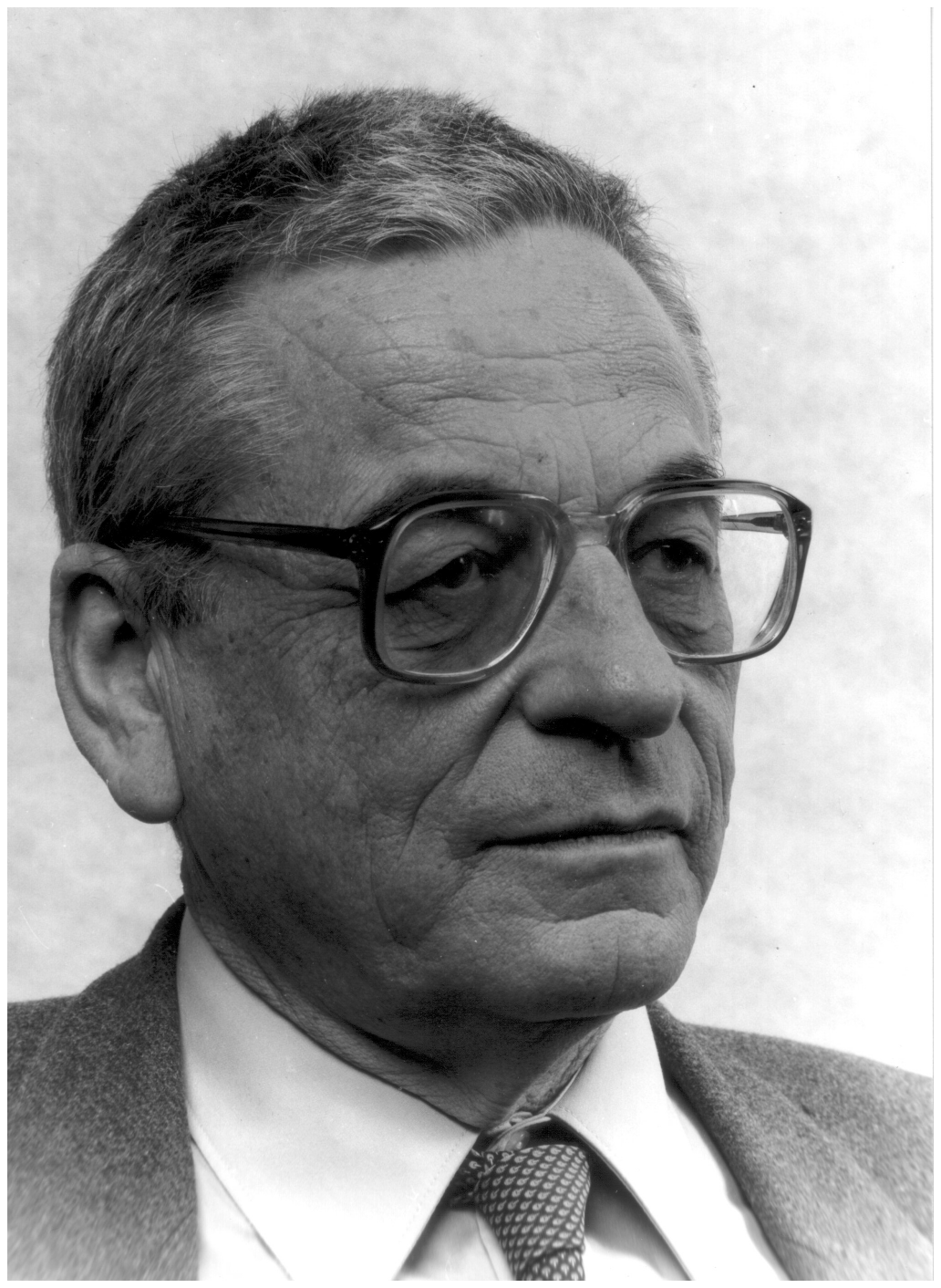}}
\caption{Nicolaas `Nico' Godfried van Kampen. Courtesy by Gerard 't Hooft}
\end{figure}
Nico kept on still going strong. In his 2008 letter, Nico writes at age 88: {\it The scandal is that there are still many articles, discussions and textbooks,
which advertise various interpretations and philosophical profundities}. So for Nico nothing changed in the two decades since his 1988 paper.
Has something changed  since then? Yes, but not only in a positive direction. 
The good news is that one class of realistic models for the dynamics of quantum measurement has been solved by
Armen Allahverdyan, Roger Balian and myself (hereafter: ABN), to be discussed below.

\subsection{The battle at the social media}

\hfill{\it Please don't bother me with your many worlds,}

\hfill{\it I have already so many problems in this world;}




\hfill{\it 
Neither bother me with your multiverse,}

\hfill{\it I have too many problems in this Universe}

\hfill{Anonymous}


\vspace{3mm}

In contrast to the situation on which Nico reflected in 2008, now there are also the social media.
One may find many clips, podcasts and panel discussions on the interpretation of QM. 
Next to giving explanation concerning interesting subjects or points of progress, they can promote subjects that would be better off with silence.
While support for science has always been based on a motivation for rewards or spin offs, 
social media emphasize one further component: popularization. Clearly, concerning the ill-informed general audience, this
``new axis in Hilbert space''  is not immediately -- and often not at all -- connected to scientific relevance.

Thus, while, for example, Sabine Hossenfelder stays on the safe side of the debate by supporting a version of the statistical interpretation,
but would be better off by not speaking about metaphysical notion of ``the wave function of the particle'', 
various others, including other respectable colleagues such as Sean Carroll and Max Tegmark, are caught by the spell of the many worlds interpretation,
and still  others, like Mario Livio, by the multiverse interpretation. 
In my opinion, Sean, Max and Mario fall in the trap of Nico's above mentioned Theorem IV.
The battle to get out of the trap has the smell of the criminal assuring his innocence.

The idea of a multiverse can neither be proven nor disproven; I do not consider this idea to be a part of science, 
since, certainly for now, there is no way to test it. Scientific guessing is needed, and some of it can become part of 
testable science, but most of it will not.
Quantum mechanics should deal with experiments in the laboratory or even in detectors in satellites, 
but it should not have to rely on what happens in the remote Universe, let alone in other universes with which no communication is possible.
Trying to convince audiences that (the measure problem\footnote{One can express the measure problem in the multiverse as:
When I do a measurement, there are fortunately other universes in which little green men with their little green women carry out the same experiment, 
helping to get my statistics right.}
 on) other universes or ``unvisited worlds'' are relevant to get the proper Born rule
for the experiments in a laboratory, seems no less than a scandal to me.
Asher Peres stressed that unperformed experiments have no results\cite{peres1978unperformed}.
There is a good reason for that: during an experiment, a (quantum) force is exerted on the particle or system;
this creates a situation different from the original one, a fact of Nature which should not be lost in vague discussion.
How this generalizes to ``unvisited worlds'' or ``unvisited universes'' is a question solely for science fiction.

So Nico's scandal is certainly not eliminated by the arrival of social media, and, let us admit, neither is it in the literature,
not even in the scientific literature.

\subsection{Praise of quantum theory}

\hfill{{\it O store gud}\footnote{How great Thou art}}

\hfill{\it O store Gud, n\"ar jag den verld beskådar}

\hfill{\it Som du har skapat med ditt allmaktsord, }

\hfill{Written by Carl Boberg}

\hfill{Sung by Sissel with the Tabernacle Choir}

\vspace{3mm}

Quantum mechanics has allowed us to understand Nature: the solid state, particle physics and even the microwave
 background of early cosmology. We are now in the era of nanotechnology,
mobile phones that do not only allow to phone everybody anywhere on the world, but also have much greater
computer power than the big computing halls that were available half a century ago.

There is no doubt about what  the quantum formalism embodies, be it that of quantum mechanics or quantum field theory.
The structure of Hilbert space and Fock space is well known. There are strict rules for the allowed operators and how to handle them. 
The Schr\"odinger equation and  the ensuing quantum field theory describe the whole condensed matter. 
The proof of renormalizability of quantum gauge theories by 't Hooft and Veltman allowed quantum field theory also to describe 
the whole particle physics. The predictions are unique, given the choice of Hamiltonian or Lagrangian. 
For a historical overview, see \cite{plotnitsky2011reasonable}.

In laboratories, repeated experiments are performed to obtain statistical accuracy so as to compare to theoretical predictions.
At the Large Hadron Collider in Geneva (LHC)  the discovery of the Higgs particle
was an exciting achievement, finishing off the standard model (SM) of particle physics. But, to the regret of many,
no further fundamental physics has been discovered so far. This has been termed ``the nightmare scenario''.
The Dutch language allows to put this in the succinct phrase ``De Higgs en verder niks''
(The Higgs and further ``nicks'').

Theoretical predictions for the occurrence rates of many interaction channels has led to an impressive confirmation of the 
SM, with mild question marks for muon related channels, that may still end up as statistical flukes. 
But the point to stress here is that there is no doubt about the applicability of quantum field theory in physics.
Moreover, its ideas have been applied, among others, by Andrei and his pals in cognitive science, psychology, 
genetics, economy, finances, and game theory alike\cite{khrennikov2014ubiquitous};
for now, the  only question is whether the SM Lagrangian is still good enough, or that lepton universality is absent.


\subsection{Is quantum theory the formalism plus the postulates?}
\label{Darrigol}

\hfill{{\it E\'en plus \'e\'en is twee}\footnote{One plus one is two}}

\hfill{Dutch expression}

\vspace{3mm}

The (Copenhagen) measurement postulates are so well known and rightfully so often taught in freshmen courses,
that they may leave the impression that quantum theory consists of two parts:
first the quantum formalism (wave functions, Hilbert space, Schr\"odinger equation, Fock space, operator formalism, \ldots)
and second, the postulates. This gives the postulates a higher status than they deserve.
They are merely shortcuts.

In a laboratory one may observe many objects: devices, machines, cables, coolers, ventilators, next to students and the supervisor.
But {\it one does not observe postulates}.  The postulates are an abstraction for what happens in a realistic measurement.
The proper physics is to model the experimental setup (``context'') and the ensuing dynamics that constitutes the measurement.
That deals with the only point of contact between the quantum formalism and the reality in the lab,
so interpretation of the formalism should be based on that, not on creative (wishful) thinking.

To describe a realistic measurement setup is a hopelessly complicated task, and that has hampered progress in quantum foundations.
The best to strive for are idealized models, which are still rich enough to capture the essential physics, but can be handled
theoretically, preferably be solved analytically. Even this was a long road, the various models have been reviewed by 
ABN\cite{allahverdyan2013understanding}.
The Curie-Weiss model for quantum measurement, to be discussed in section \ref{Qmess-sec}, serves this purpose.
Its solution is rich, with various aspects of how the dynamics of the measurement is proceeding. But often one is not interested
in the apparatus, only in its indications of the pointer (measurement outcomes). 
To a good but incomplete extent, the essentials of the indications are captured by the postulates.

\myskip{ 
The postulates are a shortcut to the dynamics of the measurement, no more, no less. 
They are a kind of black box, and as such, they are helpful in practice, but not helpful for interpretation of the quantum theory.
For that, one needs the deeper level of the description of the measurement dynamics.
That will provide various results from the formalism itself, so that less stringent and more specified postulates suffice.
} 

The principles for a general physical theory are formulated by Darrigol\cite{darrigol2015some}. They involve 4 steps:
\begin{itemize}
\item{A mathematical formalism, here the Hilbert and Fock spaces, symmetry groups, gauge structure, etc.}
\item{Physical laws, here the Schr\"odinger and Liouville-von Neumann equations, shortly: Hamiltonian dynamics}
\item{Interpretation, understood as a way to interpret some idealized experiments}
\item{Approximations. They are needed to discard irrelevant aspects so as to come to `clean' statements }
\end{itemize}
In quantum physics, where an experiment influences the system, one also needs its setting (context)\cite{auffeves2016contexts},
\begin{itemize}
\item{A module: the principle or postulate that individual (idealized) experiments yield individual pointer values  in a given experimental context}
\end{itemize}

\subsection{The ontology of Nature needs not be encoded in quantum mechanics}

\hfill{{\it Das Leben is hart, }

\hfill{{\it aber ungerecht}\footnote{Life is hard, but unjust}}

\hfill{German saying}

\vspace{3mm}

Individual experiments yield individual outcomes. One may ask the question: what happened such that this outcome arose? 
High energy cosmic ray particles hit the atmosphere now and then, and can be detected by large collective instruments such as IceCube and Auger. 
A standard question is: how did this particle arrive here? The standard answer is: it got ejected long ago in some remote stellar explosion, 
and travelled all the years with one goal, namely to hit our detector. All of this is an ontology, with the question: {\it how} did it precisely happen?
Much effort has been devoted to this question, see, e.g.,  Andrei's works \cite{khrennikov2014beyond,plotnitsky2015reality}.

Since quantum mechanics works so well in practice, it fulfils its primary  task of offering an epistemology for the experiments,
 which is good enough: quantum mechanics works.
If it would offer more, such as an ontology, that would be pleasant, but if not, so be it. 
There is no reason for a theory to be final or to deal with all aspects.
Newton's theory does not deal with the atomic structure of the planets and apples it considers,
but it is still of great value in its domain of application. Nobody will dismiss Newton theory and, likewise, there is no reason to abandon QM. 
Still, there are other physical questions, that may once be described in a theoretical framework, such as:
in a double slit experiment, through which slit went the first electron that hit the screen?

Bohmian mechanics offers an ontology, but we consider it as inapt, as we discuss in section 3.6.

\subsection{How I got here}

\hfill{\it Get here, if you can}

\hfill{Written by Brenda Gordon}

\hfill{Sung by Oleta Adams}

\hspace{3mm}

Let me recall coffee time at the Institute for Theoretical Physics at the University of Utrecht during my Ph.D.  time in the early 1980s. 
All the staff came for common coffee and discussion, and the informal Friday seminar. 
The coffee was prepared by the youngest student.
At some moment Nico started to discuss the foundations of quantum mechanics. When Nico spoke, everybody listened,
including professors Johnny Tjon in nuclear physics,  Tiny Veltman and Gerard 't Hooft in particle physics,  
the latter was my master thesis supervisor, and further my most honorable Ph.D. supervisor Theo Ruijgrok in stochastic and quantum physics,
and my later postdoctoral supervisor Matthieu Ernst in statistical physics.
 But busy with their own subjects, they all left one by one, when no clear conclusion was reached.
When the topic got stretched out over weeks and even months, 
it ended with Nico testing his ideas solely on his colleague Ben Nijboer, statistical physicist, deep thinker 
and every inch a gentleman. And I was the only other one left; I was the audience for this play without script\ldots

Ben exposed a firm opinion about doing a Ph.D.  on quantum foundations: A student should first learn how quantum mechanics works;
if he has written some 20 papers, he has worked enough with it to have acquired a solid basis for the foundations
if he is interested; before that, it can not lead to a serious contribution. 
I followed Ben's advise, for two decades I just ``had no time'' for foundations.
I carried out research on other subjects, mostly in statistical physics.
In fact, I liked Nico's courses on mathematical physics so much, that my first research decade was devoted to 
toy models of statistical physics \cite{nieuwenhuizen1983exact,de1985distribution}. 
For my field-theoretic approach to compact random walks with relatively few distinct sites visited,
 published in 1989\cite{nieuwenhuizen1989trapping} and confirmed numerically
  in $d=2$ \cite{vRossum1993band} and $d=3$ \cite{van1994density}, 
I expected it for mathematicians to take 30 years to prove the result firmly, but it appears that they need more.

It was my move to the experimental group on multiple light scattering of Ad Lagendijk in Amsterdam that taught me to look 
with the eyes of an experimenter\cite{den1993location,de1994probability}, 
and this profoundly changed my view on what is most relevant in physics,
to solve various questions in this field\cite{nieuwenhuizen1993skin,nieuwenhuizen1995intensity,van1999multiple}.
Working also on spin glasses, I found the first model where the 
Parisi function $x(q)$ for the overlaps between states is solved analytically \cite{nieuwenhuizen1995exactly}.
Remarks by Giorgio Parisi about the breakdown of the Ehrenfest relations for glasses led me to solve this 
classical paradox\cite{nieuwenhuizen1997ehrenfest}
and  next to formulate the thermodynamics of glasses \cite{nieuwenhuizen1998thermodynamics}, 
which had been a hot topic in the 1950's, 60's and 70's.
With Ph.D.  student Luca Leuzzi we wrote a book on the subject\cite{leuzzi2007thermodynamics},  
Giorgio was so kind to write the preface.

Next, I wondered what subjects to work on in the next decade.
Just around that time in early 1999,  Armen Allahverdyan joined my group and we went over many subjects
of common interest, including quantum thermodynamics and interpretation of QM.
Not long thereafter, we were joined by Roger Balian, forming the ``ABN'' team for 
our research on quantum thermodynamics; I coined the by now well known term ``ergotropy'' 
for the maximal work extractable from a quantum system\cite{allahverdyan2004maximal}.
A second theme was our ABN decades long research line on quantum measurement,
the main theme of the present essay.
What I present below is my view on our achievements and related issues.

\subsection{Disclosure}

\hfill{\it I fully take no responsibility for this}

\hfill{Anonymous}

\hspace{3mm}

While my own thinking is inspired by my teachers, Nico in particular, and even the more so by my coworkers Roger Balian and 
Armen Allahverdyan, this does not imply that I follow anyone of them blindly, and, perhaps unsurprisingly,  this happens to be completely mutual.
They bear no responsibility for the present viewpoints, but would likely do well in agreeing with all of them.
Discussions at the various V\"axj\"o conferences were very illuminating and stimulating, for which I thank the various participants
and Andrei in particular.

In section \ref{Meaningrho} we discuss some issues of the density matrix.
In section \ref{OnInterpretations} we discuss some interpretations of QM.
In section \ref{Qmess-sec} we analyze a model for ideal quantum measurement.
In section \ref{sinequanon} we stress the need of models for quantum measurement for the interpretation
of the quantum formalism.
Finally, in section \ref{elephant} we stress the need for an ontology.

\section{Meaning of the density matrix}
\label{Meaningrho}

\hfill{\it Say you don't mind}

\hfill{Written by Denny Laine}

\hfill{Sung by Colin Blunstone}

\vspace{2mm}

The interpretation of ``the wave function'', or, more generally, the density matrix, has been subject of
fierce debates in last century.

\subsection{The least biased interpretation of the density matrix}

The least biased interpretation of the density matrix in quantum mechanics is that it codes, in an abstract manner,
our best knowledge about the ensemble of identically prepared systems.
The ensemble can be a beam of neutrons during one day, or the collisions in the proton beams at the Large Hadron Collider at
the CERN during one season. But it may also refer to one ion in a Pauli trap, which is repeatedly excited by a lasar beam
and repeatedly emits a photon. Finally, for macroscopic systems, one may imagine a Gedanken ensemble of similar systems. 

Like Gibbs state in classical statistical mechanics, the density matrix of an ensemble of quantum systems is a 
catalog of incomplete knowledge. In both cases, the entropy is a measure for the lack of knowledge.
So let me recall the modern view on entropy.

\subsection {Entropy in classical statistical physics}

\hfill{\it Increasing knowledge,}

\hfill{\it  decreasing satisfaction} 

\hfill{Anonymous}

\hspace{3mm}

In Newtonian dynamics of a large set of particles, the dynamics can be followed for some time.
It can be said that the entropy is zero then (or, at least, conserved in time). Indeed, entropy is a measure
for the lack of information, but in that early time window we know the positions and speeds of the particles,
so no information is lost. But at larger times an exponential divergence between orbits starting close to each 
other, set by the Lyapunov  rate (unfortunately called ``Lyapunov entropy''),  is unavoidable due to chaotic behavior over long times.  
Weather forecasts, however, bear more upon chaos theory (since the 1970s) than upon classical statistical physics.
It is the very reason why reliable forecasts are still limited to a few days despite the enormous  increase in computer 
power in the last decades, with only a modest improvement in accuracy.

In classical statistical physics  the modern view is that entropy codes our lack of knowledge of or the missing information about
the physics at the microscopic scale\cite{jaynes2003probability,balian2004entropy,balian2005information,brillouin2014scientific}.

Actually, there are many ways of coarse graining and there are many entropies. The founding fathers of thermodynamics already
introduced the Clausius entropy, the thermodynamic entropy and the Boltzmann entropy. 
They considered macroscopic systems, so that the leading term
in the volume coincides, which allows to speak about ``the entropy''. However, if the system is small, 
while both the bath and the work force are large, one enters the field of Quantum Thermodynamics. 
Then there are many entropies and they are all different.  Each of them relates in principle to a measurement which could 
be performed on the system. Hence entropy is a state of knowledge, defined in an experimental context, true or in Gedanken.

The ``state of knowledge'' role is exemplified by the Gibbs paradox\cite{gibbs1906scientific}. Imagine a  pool table with many 
green and blue billiard balls in some random configuration.
A fully sighted observer will associate a certain entropy to it. But a color blind observer will only see one kind of balls 
and add an additional ``mixing entropy'', $k_B\ln 2$ per ball.
Gibbs himself considered two identical compartments with a gas; removing the wall between them, also leads to this mixing entropy.
Taking the Gibbs paradox into quantum thermodynamics, Armen and I showed that the difference in entropy can
be connected to the possibility of extracting work from the setting difference, which depends on the abilities (``knowledge'') of the experimenter
(``observer'') \cite{allahverdyan2006explanation}.

Somewhat related to this is the demonstration that independent (tensor product of) quantum states of identical particles acquire their
symmetrization dynamically in their collision \cite{allahverdyan2021dynamical}. So the symmetrized wave functions always used in the solid state
need not be used when considering electrons partly at the Earth and partly at the Moon.

\subsection{The density matrix for microscopic systems}

\hfill{{\it Veel weten van een klein beetje}\footnote{Knowing much of a little bit}}

\hfill{Dutch expression}

\vspace{3mm}

A similar observer-dependence applies to the meaning of the quantum wave function or, more generally, the density matrix.
The density matrix is an abstract mathematical object that describes our best knowledge of the considered ensemble of identically prepared systems. 
Notice that it describes {\it our} best knowledge, to be employed for prediction of outcomes of future experiments.

Concerning ``best knowledge'', the following example was considered in footnote 29 of our ``Opusculo'' \cite{allahverdyan2017sub}: 
A spin $\half$ is measured along a certain axis.
Observer A knows the axis direction and the results, so he can construct the collapsed post-measurement state.
Observer B knows that the measurement has taken place, but not the direction of the measurement axis.
The best he can do, is to average the final density matrix over all directions of the axis.
We may add to this Observer C, who is unaware of this measurement and will continue with the original density matrix;
he has least (namely: no) knowledge of the measurement. The best statistical predictions for future experiments can be made by
Observer A, the second-best by Observer B and the least-best by Observer C.

\subsection{Contra ``the wave function of the particle''}

\hfill{\it The wave function of the system}

\hfill{\it is a misconception;}

\hfill{\it The wave function of the Universe}

\hfill{\it is a cosmic misconception}

\hfill{Anonymous}


\vspace{3mm}

A person is not a wave, also not when taking part in a wave in a sport stadium.
The wave is a description of the collection (ensemble) of persons. A given person can participate in many different waves, 
so it is quite a stretch to proclaim that the wave is some type of a property of a person in it.
Moreover, the behavior of any person will depend on the environment he/she is in; 
I find it amusing that this sociological property has an analogon in quantum physics, or perhaps even its root in it.

In my opinion the very same holds for ensembles of particles, like the beam of protons in the LHC beam during some period.
The same protons could have been a member of a different beam, at another moment, hence to me 
``the wave function of the particle'' is a horribility. Setting apart my opinion, this concept is not needed anyhow, and actually only a burden.
Employment of Ockham's razor allows a cleaner scientific situation.

\subsection{Contra ``the particle is here and there at the same time''}

\hfill{\it The electrons in my pinkie are mine!}

\hfill{Anonymous}

\vspace{3mm}

In colloquial presentations, often statements are made like  ``in a Stern--Gerlach experiment, the electron  is at two places at the same time''.
This is an absurdity, a nonsense, a scandal. No measurement has ever been done to support that claim, and, actually, 
none can ever be done. It is a colloquial way of trying to express some of the wonders of quantum mechanics,
in particular the wave nature. But Einstein said already: one must simplify as much as possible, but not further.
In the end, quantum mechanics will remain a mysterious theory.

The culmination of this reasoning was proposed by John Wheeler in a telephone call to Richard Feynman in the spring of 1940, 
``all electrons are the same, actually there is only one electron popping up here and there''\cite{wikiWheeler}. 
Needless to say that this absurdity (scandal) causes questions in electromagnetism and gravitation.
To stress that a sound approach is possible, we recall the treatment of Allahverdyan et al,
in which electrons at a certain location in the initial state are treated as really different from the ones at other locations 
by the (non-symmetrized)  tensor product\cite{allahverdyan2021dynamical}.

\subsection{Pure state as a purified limit of a mixed state}

\hfill{{\it Dat is geen zuivere koffie}\footnote{That is no pure coffee}}

\hfill{Dutch saying}

\vspace{3mm}

A pure state $|\psi_n\rangle$ corresponds to a density matrix $|\psi_n\rangle\langle\psi_n|$, while a mixed state involves a mixture of them,
 $\sum_n p_n|\psi_n\rangle\langle\psi_n|$, with $p_n\ge0$ and $\sum_np_n=1$. It is often thought that a pure state has a special meaning, 
 like being attachable to a particle, but we oppose this point of view. In a statistical approach, a pure state is a limit of a mixed state,
 where all $p_n\to0$, except for one, which goes to 1. No special meaning is gained in this limit: also the resulting pure state stands 
 for an ensemble of identically prepared particles or systems.

Since the general case is a mixed state, it has to be purified to reach a pure state.
To purify one variable, one performs a measurement, selects the members with the desired outcome, and discards the other members.
Then a second variable, commuting with the first, can be purified, by a similar procedure; next a third one, commuting with 
both previous ones, can be purified, and so on. However, in practice only a handful (or perhaps a dozen) of purifications can be carried out. 
For one reason, because of the difficulty to do any experiment, for another, because one would have to discard nearly all members 
of the initial ensemble. This is uneconomic, to say the least, and an over focus on very special cases.

\subsection{Schr\"odinger's ambiguity}

\hfill{\it In psychology, }

\hfill{\it the opposite is also true}

\hfill{A. Ross}

\vspace{3mm}

The quantum formalism has an unexpected property. Suppose we have one purified beam of neutrons with spins in the $+z$ direction and a similar one
with spins in the $-z$ direction. Combining them yields the mixed density matrix 
\BEQ
\half\big(\, |\uparrow\rangle\langle\uparrow| +|\downarrow\rangle\langle\downarrow| \,\big)=\half \begin{pmatrix}1&0\\0&1\end{pmatrix} .
\EEQ
However, the same result is obtained when combining similar beams in the $\pm x$ directions, viz. 
\BEQ
\half\big(\,|\rightarrow\rangle\langle\rightarrow| +|\leftarrow\rangle\langle\leftarrow|  \,\big)=\half \begin{pmatrix}1&0\\0&1\end{pmatrix},
\EEQ
or by combining similar beams polarized in arbitrary $\pm \,\hat n$ direction. 
This implies that after the merging, no test can be performed to establish which beams were combined.
This simple example shows that a mixed density matrix can in general not be related to individual systems. 
It is a form of the measurement problem, that is, the inability of QM to describe individual events,  i.e., the lack of an ontology of quantum theory.
Since in our view a pure state is a limit of a mixed state (obtained after purification by a set of measurements), it does not obtain
a special meaning in this (never really reached) limit, so it can neither be connected to individual particles.

\subsection{Individuality of macroscopic systems}
\label{IndivMacro}

\hfill{\it Me, myself and I}

\hfill{Expression}

\vspace{3mm}

Microscopic systems, like atoms or cosmic ray particles, exist, but there is presently no tool to describe them 
other than classically. Quantum theory captures their behavior in a statistical way.

Classical systems are macroscopic, they typically have some $10^{25}$ atoms and degrees of freedom.
One can identify a few of them as ``system properties'', like point of gravity, the speed, the temperature\ldots .
From the abstract quantum formalism point of view, it is natural to postulate that when a density matrix contains a mixture of such macroscopic states,
it can be decomposed in {\it individual} macroscopic states. Along this line of reasoning, individuality is a phenomenon emerging in the thermodynamic limit,
and overcoming Schr\"odingers ambiguity. Still, this emergence likely only holds, and is only needed, for the ``classical'' system variables: 
``what is quantum should remain quantum''.

While needed for interpreting the final state of a quantum measurement, this principle is much more general. 
Imagine two dynamical descriptions of an Ising magnet in zero field below the thermodynamic phase transition temperature. 
In the first approach, one may work with classical spins $\pm1$ and employ Glauber dynamics, which, because of the classicality, 
deals with individual systems. Alternatively, by directly employing the quantum approach with its quantum dynamics, 
one may follow Darrigol and idealize the situation\cite{darrigol2015some} by neglecting the anyhow tiny  ``cat'' terms, 
after which the classicality postulate allows to connect to individual systems.

In the alternative view that the quantum formalism is just a technical framework to describe outcomes of measurements, one starts 
from the pointer readings on a macroscopic apparatus, and the postulate is evidently not needed 
\cite{de2014quantum,de2015quantum,auffeves2016contexts,auffeves2020deriving}.
Another treatment of the context is by Andrei \cite{khrennikov2003contextual}.

\subsection{Classical systems are mostly quantum and thermodynamic}

\hfill{\it Je kunt niet verder springen}

\hfill{\it  dan je polsstok lang is\footnote{You can't jump further than your vaulting pole allows}}

\hfill{Dutch saying}

\vspace{3mm}

The notion of a ``classical apparatus'', so often discussed in connection to quantum measurement, is an oversimplification. 
What the founding fathers like Heisenberg and Bohr were aiming at, 
was, of course,  the pointer of the apparatus. In order that it can be read off or processed electronically, the pointer has to be macroscopic
and thus have classical features, but its dynamics will involve, in principle,  all the microscopic degrees of freedom.

Like every system in Nature, an apparatus consists of atoms, and many of them, say, $10^{25}$, and it has even more degrees of freedom.
This holds also for any ``classical'' system. We can speak about classicality of a few of its macroscopic properties,
like the position, momentum, angular momentum and temperature. In a fluid one often considers local quantities like density, pressure, temperature, 
average speed (wind) and so on, which are averages in a mesoscopic cell, still containing some $10^{12}$ particles or so.
But even here the mentioned ones are only a handful out of the  $10^{12}$ degrees of freedom.
Nearly 100\% of the degrees of freedom remain quantum and never become classical.

As an example, in  section 10.1.2 of their ``Opus Magnum'' \cite{allahverdyan2013understanding}, ABN consider a small metallic grain. 
Its center of mass can be treated as a classical variable, but at low and moderate temperatures  its specific heat 
must be described by Debye theory, that is to say, by quantum mechanics.
So Heisenberg's celebrated classical-quantum cut depends not only on the considered system or variable
but even on the condition; the classical regime of the specific heat of the grain pertains to high temperature.

\subsection{No apparatus is microscopic}




\hfill{\it An elephant cannot easily hide itself}

\hfill{Expression}

\vspace{3mm}

A ``microscopic apparatus'' is a contradictio in terminis; it is not an apparatus, but just another quantum system.
To say that it ``measures'' something is misleading (it is a scandal);
like any other quantum system,  a ``microscopic apparatus'' just undergoes some quantum dynamics,
that can be tested by employing a macroscopic apparatus, of which the outcomes can be read off or processed.

 It is often said that the Stern-Gerlach magnet acts as an apparatus. This is misleading (thus another scandal),
the apparatus is the screen that collects the electrons in the Stern-Gerlach setup.
The also popular term ``pre-measurement'' does more justice to the action of the magnet.\footnote{Since the divergence of the 
magnetic field is zero, a gradient in $B_z$ has to be accompanied by a gradient in $B_x$ and or $B_y$,
which cause non-idealities. Hence no Stern-Gerlach 
setup is ideal; an ideal Stern-Gerlach setup is the mathematical limit of a series of setups with ever larger Stern-Gerlach magnets,
which involve ever smaller gradients in the magnetic field.}

One often sees the apparatus as having both quantum and classical parts with the outcome of the interaction (the registration of the measurement), manifested in the classical part; however, as is evident from ABN's solution of the Curie-Weiss model, this ``classical part'' has
a dynamics that stems purely from the quantum nature of its constituents. 

In fact, quantum objects can be macroscopic, too  (Josephson junctions) even though their quantum nature is defined by their microscopic constitutions and can only be detected by a suitable apparatus. 

Not every system can act as an apparatus. Obviously, 
 a broken apparatus, or even an apparatus not  under the right conditions, does not act as a proper apparatus,
let alone other objects, such as rocks. These trivial examples corroborate that a working apparatus is a special macroscopic system
under special conditions, adept for the task under consideration.

\subsection{No macroscopic system, cat nor Universe, can be in a pure state}

\hfill{\it Next time I see a Schr\"odinger's cat,}

\hfill{\it I won't be able to contain myself}

\hfill{Anonymous}

\vspace{3mm}

The idea that a cat or an apparatus, or for the sake of the matter, any macroscopic system, can start in a pure state 
(on the eigenbasis of some operator), is just another scandal.  A state is a mathematical description of the knowledge
about the ensemble of identically prepared systems, in reality or in Gedanken.
For a macroscopic system, only a relatively very limited amount of information can be known and described.

Starting from a complete description, going to a reduced one amounts technically to tracing out all other degrees of freedom.
 In the Liouville-von Neumann equation of motion for the density matrix $\rho$,
viz. $\d\rho/\d t=(i/\hbar)[H,\rho]$, the tracing is easily carried out on the left hand side. The right hand side involves
the trace of the commutator. Except for extremely fine-tuned Hamiltonians, no simplification occurs, 
so that the traced equation is not closed. It can not be solved and at best be used for very special purposes or in approximation.
The physical reason herefore is, of course, that the multitude of microscopic degrees  of freedom can not be neglected; they are all relevant,
and in principle they all influence the dynamics of reduced system parts such as the pointer. 

Adherents of many worlds theories or spontaneous collapse models omit this check, so that they effectively work
with extremely fine tuned  but unspecified Hamiltonians, or, more often, abandon modelling at all, 
and restrict themselves to arguing with linguistic concepts, not guided nor constrained by detailed analysis of the physics.

Hence a {\it cat} can not be in a pure state. The statement  ``the cat is in a superposition of a state {\rm dead} and a state {\rm alive}"
is a poetic construct with no bearing on reality. The conclusions drawn from it already convinced Schr\"odinger 
that this line of reasoning leads to absurdity (to scandals). We side with him.

An {\it apparatus} has to be macroscopic, so it can not be in a pure state. 
This invalidates the von Neumann-Wigner theory for quantum measurements and exposes the Many Worlds Interpretation as inapt. 
Instead, in a quantum measurement with the apparatus necessarily in a mixed state, the total system of tested system 
plus apparatus always starts out in a mixed state, also when the tested system starts out in a pure state. Consequently,
there is no unitarity paradox: the system itself can go from pure to mixed, since during the measurement it is an open system coupled
to the apparatus. 
More precisely, the unitarity paradox is solved by discarding the off-diagonal blocks,  after they have become irrelevant
for all practical purposes, in line with step 4 in Darrigol's general scheme of section \ref{Darrigol}.


A {\it black hole} can not be in a pure state, certainly not an astrophysical one that starts out from an imploding star, but neither one
that arises by the collapse of some chunk of matter in the early Universe.
Hence there is no black hole unitarity paradox; Hawking evaporation will not cause a transition from a pure state to a mixed one
because considering a pure state is a mathematical or philosophical exercise unrelated to  a physical black hole.
The fine grained entropy (von Neumann entropy) is conserved under unitary motion, also
when Hawking radiation occurs. It is impossible to measure the fine grained entropy in the solid state,
say when burning coal, let alone by a Dyson sphere of detectors around the black hole.
The coarse grained entropy of a closed system increases
according to the second law of thermodynamics. The increase of coarse grained entropy in Hawking radiation 
is compatible with the decrease the coarse grained entropy of the black hole, as only the sum has to increase.
It can be summarized as ``the information goes out with the Hawking radiation'', and ``finally the black hole has very small entropy''.

The {\it Universe} can not start out in a pure quantum state. Quantum theory describes our probabilistic knowledge about the 
Gedanken ensemble we imagine it to be a member of. No set of apparatuses for measuring commuting variables can be imagined 
which would purify the mixed state describing our incomplete knowledge about an ensemble of identically prepared Universes. 
And it seems difficult to give a meaning to discarding the very, very nearly 100\% of the candidate universes with wrong measurement outcomes.
Assuming to have the absolute knowledge about (the ensemble to which) the whole Universe (belongs), 
is making oneself equal to God. I would beg: then please make next step, and also create, for instance, another Universe.
Unitarity considerations (to go from a pure to a mixed state in a measurement) 
based on ``the wave function of the Universe" are mental exercises that have no bearing on our Universe; they can be called ``poetry''.
Even so, this is not to claim that applications of e.g. the Hartle-Hawking wave function will always lead to false results, 
but to stress that its use should be motivated.

\subsection{Many mathematical issues are devoid of physical relevance}

\hfill{\it Much ado about nothing}

\hfill{Shakespeare}

\vspace{3mm}

The von Neumann entropy is conserved in time and so is the trace of any power of the density matrix.
These quantities express the unitary character of the evolution, but they can not be measured and there is nothing physical that can be deduced from them.
 
It is possible to consider the wave function of a cat or of the Universe, thereby imagining that the mixed density matrix of such
entities has been purified completely. However, there is no sound way of even imagining that the purification can be carried out in any reality.
Hence in my opinion there is no sense in considering such concepts, rather they are a reliable source for paradoxes.
Not a complete ridicule.
A more physical and urgent question seems to me: how many angels can dance on the tip of a needle?

Whether or in how far the landscape of string theory and the many universes in the multiverse fall in this category, 
is an interesting point for reflection.

\section{On some interpretations of quantum mechanics}
\label{OnInterpretations}

\hfill{\it For progress in this field}

\hfill{ \it less study is needed}

\hfill{Anonymous}

\vspace{3mm}

The interpretation of QM has eluded the community for long. 
Erwin Schr\"odinger originally supposed that his wave function describes the charge density, but Max Born brought the connection to probabilities,
which is so deep, that now, about a century later, it is still debated. Historical papers on the subject are collected by Wheeler and Zurek  \cite{wheeler2014quantum}.

\subsection{Interpretatio Copenhagenensis}

Most often employed is the so-called Copenhagen interpretation. It serves in introductory courses on QM,
since a short cut for interpretation will work for the many applications. It is also often employed in practice
where one needs to compare measured frequencies with theoretical probabilities, it is  David Mermin's minimalistic and empirical 
``shut up and calculate'' approach without bothering about the pitfalls of the foundations.

While there is a broad literature on its various formulations, we recall the main points: 
QM applies to individual objects in a probabilistic way; a particle has its wave function (wave packet);
observation leads to reduction of the wave packet (collapse of the wave function); the Born rule holds for probabilities of outcomes;
and the quantum description is independent of the mind of the experimenter.

As seen in section 5, 
ABN basically support this interpretation, except for their more modest interpretation of the wave function or density matrix.

\subsection{Contra multos mundos}

\hfill{\it  To excel  in explaining concepts}

\hfill{\it that cannot be understood}

\hfill{Anonymous}

\vspace{3mm}

After Bohr and Heisenberg presented initial ideas, von Neumann presented a first model for quantum measurement, 
so it became possible to connect interpretation to models of measurement. The ``many worlds'' interpretation (MWI) 
starts out with supposing that the system S and the apparatus A both begin in a pure state. 
By the coupling, A is said to evolve to a pure state correlated to S. But this is a magic dynamics, if not a Kindergarten dynamics.
In Nico's terms: it is a scandal that it is still presented.
The MWI has never been backed up by a solvable  model and, let us admit, it is counterintuitive and not economical.
The bishop of Ockham would find a worlds-wide if not heavens-wide task here for employing his razor.

 The MWI is sometimes termed the ``many {\it words} interpretation'', and there is some truth in this bizarre and even ridiculous term, 
 which also entered the title of this work. Indeed, many papers have been written
 where authors try to explain various issues in sharp words, and the internet is flooded with discussions about it.
 That something can not be explained, as is claimed, has indeed a spell for many people. 
 While society could choose to go to the next level and decide to any answer scientific question 
 by a popularity poll rather than by fundamental research, 
 that would obviously be a dead end street for the development of new materials and devices. 
 ``QAnon'' physics may play a political card, but it can not bring long term progress and prosperity for society.
 Nature is just as it is. If it is to be liked, we better adjust our taste to it. Personally, I have a sweet spot for things that exist 
 and mechanisms that work.
 
 The problem with approaches without stringent mathematics is that there are far more concepts than words. 
 Not only for the Bible can one argue for hours about the meaning of a certain phrase,
 it can and has also be done for statements by physicist and philosophers alike, including famous people like Einstein, Bohr and Feynman. 
 Not intuitive reasoning, but only a stringent handling of the well determined mathematical 
 structure of QM can offer the hope of reaching a sensible and, hopefully, unique interpretation of QM.
   The unease of this field is summarized in David Mermin's mantra ``shut up and calculate'',  
 don't  waste your time on this philosophical issue, but work on verifiable predictions.
After all, ``the'' Copenhagen interpretation works quite well in practice.

 After physicists gave up the hope of ``solving the measurement problem'', it was left for philosophers.
 They did not solve it either. In hindsight, this is not surprising since for long the physicists had no sufficiently good models to direct them. 
 Working from the postulates is just a black-box treatment of measurements, which bypasses a lot of physics
 that is covered in the dynamical solution of an ideal measurement.
But in a laboratory, a measurement is the bread and butter of the experimenter, rather than a black box.
Philosophy cannot bypass a physical process, it should, instead, closely follow what happens in the laboratory.

In the many-worlds interpretation, reduction of the density matrix is even denied, and regarded as a delusion due to the limitations 
of the human mind.  Yes, it is often proclaimed that our minds are not evolved to understand certain abstract notions. 
But that smells like another scandal, ``a proof by complete intimidation'',
to cover up the fact that the proper framework or  tool for the question under consideration has just not been discovered.

Below we discuss that the truncation of the density matrix on the measurement basis does occur in a fundamental analysis of 
 dynamics of a quantum measurement.  Hence the MWI proponents should now take up the challenge of demonstrating that 
the MWI dynamics is also derivable in a serious microscopic model, to avoid scandal. Or else, to give up on it.
 
 The failure of MWI can be understood as follows. Often, the approach is to reason from imagining to follow a particle along its path 
 in time, during which measurements are performed. The assumption -- even by many though not all adherents of the Copenhagen interpretation --
 that the particle has  its own wave function is particularly unfortunate here.  At each measurement, 
 its wave function is said to split up so as to create  one new world; this idea is as mind boggling 
 as science fiction and equally absurd;  it is a scandalous story told  about the particle based on no more than wishful thinking, 
 more precisely, on the hope that QM embodies an ontology. Which, alas, it does not.

\subsection{Contra theoriam Bohmi}

The double slit experiment can be carried out at such a low intensity, that the probability for two electrons being present
at the same time, is very small. While the resulting many-electron interference pattern on the photographic plate 
will not be different from setups with higher intensity where the electrons may overlap in time, 
this clarifies one issue: electron-electron interaction does not lie at the heart of quantum interference
and the Bohm potential.

The de Broglie-Bohm theory (Bohmian mechanics) arises from the map of the complex wave function of a particle to its amplitude and phase,
after which the amplitude is connected to the probability density of the particle and the phase to its velocity\cite{bohm1952suggested,holland1995quantum}.
It is supposed to be an ontology of the particles involved. They move in a potential which is the sum of the external potential and a ``quantum'' potential,
set by the shape of the wave function. But  the low-intensity experiment dashes the hope that Bohm's ``quantum potential'' can be the cause of 
quantum interference.

While it is intuitively clear to view the Bohm trajectories as real particle trajectories, there is another problem: measurement.
One can not just read off the coordinate of the particle and be allowed to say that the particle is at that point.
Since Bohm's theory makes the same predictions as standard QM, the dynamics in the Curie-Weiss model for quantum measurement, should 
be expressed in Bohm's framework. Hence the already complicated dynamics must be written in a much more
complicated way, with much pain and without any gain.

\subsection{Contra verbos Zureki}

Another issue is decoherence, important for describing the classical world from quantum mechanics\cite{zeh1970interpretation,joos2013decoherence}.
But it is a far stretch to hold decoherence as the main actor in measurement.
Although the ideas of decoherence and pointer states is sensible, 
it is by far not enough to describe quantum measurements.

In a rather abstract approach, Zurek introduces ``pointer states of the environment'', which have a long lifetime, 
and are claimed to explain quantum measurements \cite{zurek2003decoherence}. But this linguistic stretch is a scandal.
While it is true that the environment plays some role, this can not be connected to the complete measurement;
in fact, the environment is by assumption that part of our full system that we can not control.
The ``pointer states of the environment'' can not be read off, and this name is only creating confusion.
An analogy: The air that surrounds me measures my size, but this cannot be called a measurement; no one can use it to know my size.
Otherwise, zillions of solar neutrinos pass through our bodies, but they measure nothing; would they, then they would destroy life itself.

Despite their name, these ``pointer states'' have no bearing on realistic pointers that are necessarily macroscopic.
No experimenter will count on uncontrollable resonances in the environment of his setup; to the opposite, he would discard 
events where they occur as flukes. Rather, he needs a controllable setup where he can start and finish the measurement 
at his planning, and he needs a pointer that can be read off his pointer in a time window of his choice. 
To achieve this, the pointer must be macroscopic, so he needs  a macroscopic apparatus consisting of very many 
(up to $10^{26}$) atoms and an even larger number of degrees of freedom. 
When the pointer has a small number of stable states (2 in the Curie-Weiss model), one may speak of pointer states;
they are thermodynamic states of the apparatus in various models for ideal measurement, not pure states of the environment.

In the Curie-Weiss model then bath (environment) plays two roles. In the off-diagonal elements  it leads to decoherence after the dephasing 
has hidden the signal. In the diagonal elements the thermal bath acts as usual, and absorbs the free energy dumped
during the thermodynamic phase transition from paramagnet to one of the ferromagnetic states; more precisely,
it determines the form of the evolution of the magnetization distribution $P(m,t)$. 
The essential result, a magnetized final state that can be read off, is universal, however, for all good apparatuses.

We recall that a ``microscopic apparatus'' is a contradictio in terminis, because it can not be read off;
rather it is another quantum system that interacts with the tested one.

\section{Basics of the dynamics of an ideal quantum measurement}
\label{Qmess-sec}





\hfill{\it The locomotion}

\hfill{Written by King and Goffin}

\hfill{Sung by Little Eva}

\hspace{3mm}

When an experimenter employs an apparatus, he has been sure that it has a pointer, so that he can read off the result
of each measurement. Clearly, the pointer has to be macroscopic and with it the apparatus.
The operator to be measured should enter the system-apparatus coupling linearly.
In an ideal measurement, it will not change during the measurement, and  one will only ``measure what is there''.

The apparatus has to start in a metastable ``ready'' state, a thermodynamic state triggered by the coupling to the tested system to go to one of its
 thermodynamically stable states. This is also what happens in night vision: a handful of photons trigger an avalanche of electrons in the retina,
 and the generated electric current is processed in the visual cortex. Recharging is needed.

\subsection{Introducing the Curie-Weiss model for quantum measurement}

\hfill{\it Listen, people,}

\hfill{\it to what, I say}

\hfill{Written by Graham Gouldman}

\hfill{Performed by Herman's Hermits}

\vspace{3mm}

The so called Curie-Weiss model for quantum measurement was introduced and partly solved by ABN 
in 2003 \cite{allahverdyan2003curie,allahverdyan2003thequantum} and 2004 \cite{allahverdyan2004dynamics}.
Some progress reports were published as contributed papers to the V\"axj\"o meetings \cite{allahverdyan2005quantum,allahverdyan2006phase}.
In 2013 we present an overview of such dynamical models and then go on to solve many aspects of the Curie-Weiss model in over 550 
equations \footnote{I mention that number of equations to contrast this fundamental study to typical works in the field 
of quantum measurement with, at best, a few equations, often even only posing expressions of a supposed dynamics.} \cite{allahverdyan2013understanding}. 
This allows to base the interpretation on a solid basis.
 In 2017 ABN extend the approach to a class of models and pay further attention to interpretation \cite{allahverdyan2017sub}. 
 Lecture notes are presented in 2014 \cite{nieuwenhuizen2014lectures}. 
 A paper entitled ``How to teach quantum measurement'' is underway.

The tested system $S$ consists just of a single spin $\half$ and its tested operator $\hat S$ is $\hat s_z$, 
taking eigenvalues $\pm1$ in units of $\half\hbar$.
The apparatus A consists of a magnet M and a thermal bath B. M consists of $N$ spins $\half$, denoted as  $\hat {\mathbf \sigma}^{(1)}, \cdots, \hat\sigma^{(N)}$.
In the simplest model, quartets of spins have a ferromagnetic coupling energy $J$.
B is a phonon bath coupled to each spin component $\hat\sigma_a^{(j)}$ for $a=x,y,z$ and $j=1,2,\cdots , N$;
it is characterized by its temperature $T$ and Debeye cutoff frequency $\Gamma$. 
For simplicity, the bath is assumed to act independently for each $j$ and $a$.
Its coupling to M is weak and characterized by the small, dimensionless coupling constant $\gamma$. 
The setup is schematized in Figure 3.

 \begin{figure}\label{figLoic1} 
\centerline{ \includegraphics[width=8cm]{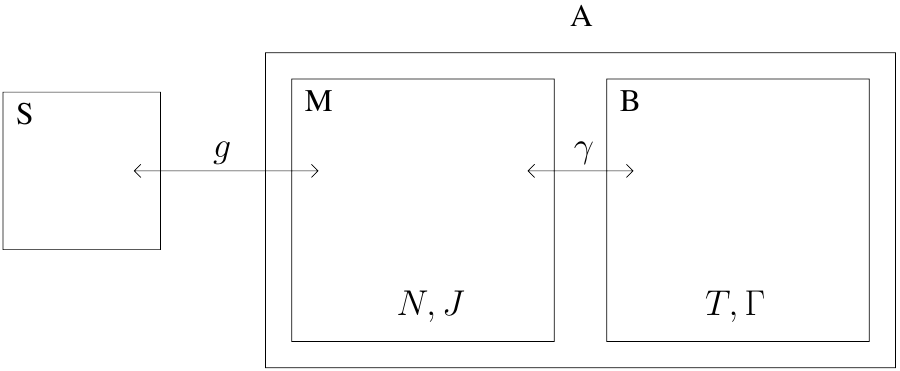}}
\caption{Schematic presentation of the Curie-Weiss measurement model and its parameters.
 The system S consists of a spin $\hat{\bf s}$.
The apparatus A includes a magnet M and a bath B. The magnet, which acts as a pointer, consists
of $N$ spins coupled in quartets through an Ising interaction $J$. The phonon bath B is characterized
by its temperature $T$ and a Debye cutoff $\Gamma$. It interacts with M through a spin-boson coupling 
$\gamma$. The measurement is started by establishing a finite interaction $g$ between the observable $\hat s_z$ 
and the magnetization $\hat M$ of the pointer; it is ended by putting $g\to 0$ after the dynamics is stabilized.
}
\end{figure}

Let the tested system S have density matrix $\hat r$, the apparatus A have a density matrix $\hat\cR$, and the total system a density 
matrix $\hat\cD$.  Initially, S and A are uncorrelated, so $\hat\cD(0)=\hat r(0)\otimes\hat \cR (0)$.
After  coupling S to A at time $t=0$, they become correlated.
At later times, the density matrix of S is defined by $\hat r(t)=\tr_A\hat \cD(t)$ and the one of A by $\hat \cR (t)=\tr_S\hat \cD(t)$. 
 The Hamiltonian of the total system has terms from system and apparatus alone,  and from their coupling, $H=H_S+H_A+H_{SA}$. 
 When measuring a system operator $\hat S$, it has to be coupled to a property $\hat M=N\hat m$ of the apparatus.
In the Curie--Weiss  model $\hat S=\hat s_z$ is the $z$-component of a single spin and $\hat M=\sum _{j=1}^N\hat \sigma_z^{(j)}$ 
the magnetization of the apparatus in the $z$-direction, 
$N$ is the number of particles in the magnet and  $\hat m$ the magnetization per particle
in the $z$-direction. The interaction Hamiltonian $H_{SA}=-g\hat S\hat M$ evidently aims to produce effects dependent on the state of S. 
The S-A coupling $g$ vanishes at $t<0$, is turned on to a finite value at $t=0$ and is put back to zero at the final time $t_f$ 
in order to decouple S from A.

\subsection{Born probabilities from a conservation law in the system-apparatus dynamics}

\hfill{\it Saved by the bell}

\hfill{Robin Gibb}

\vspace{3mm}

The greatest predictive power of quantum theory lies in the Born rule. 
In models  it is often derived by posing appropriate assumptions, see, e.g., \cite{khrennikov2008born}. 
But in a microscopic modelling of system and apparatus,
it automatically emerges from a conservation law in the dynamics  \cite{allahverdyan2003curie}, as we now discuss.

In an ideal measurement $\hat S$ will not change by itself during the measurement, so that $[\hat S, \hat H_S]=0$, or, at least, 
it vanishes effectively. Also $[\hat S, \hat H_A]=0$ because for an unbiased measurement, the apparatus is not correlated to the system. 
Hence $\hat S$ commutes with the total Hamiltonian and is conserved in the course of time. In the density matrix of S,
\BEQ
\hat r=\begin{pmatrix} r_{\uparrow\uparrow} & r_{\uparrow\downarrow} \\
r_{ \downarrow\uparrow}& r_{\downarrow\downarrow}\end{pmatrix} ,\quad 
r_{\uparrow\uparrow} + r_{\downarrow\downarrow}=1,\quad &
r_{ \downarrow\uparrow}=& r_{\uparrow\downarrow}^\ast ,
\EEQ
the diagonal elements  $r_{\uparrow\uparrow} $ and $ r_{\downarrow\downarrow}$ therefore keep their initial values, 
so that at the end of the measurement they are given by the values before the measurement, which is the Born rule.

In reality experiments are not ideal, and actually even generally not in models.
In the ABN ``Opus Magnum'' \cite{allahverdyan2013understanding} various non-idealities are considered.
For instance, a small transversal field will cause dynamics of the tested spin during the measurement process,
and cause pointer reading statistics deviating from Born's rule.
Another case is the situation where the number of spins in the apparatus is large but not very large.
Then the pointer may spontaneously ``jump'' between the upwards and downwards
magnetized state on a timescale not very much larger than the read-out time.
Also then non-idealities in the pointer readings occur, here even despite the conservation of the diagonal elements of
the density matrix of the tested system.

\subsection{Truncation of the density matrix of the system by dephasing}

At very short times, the dynamics is solely driven by the interaction Hamiltonian  \cite{allahverdyan2013understanding}. 
Since the SA coupling $-g\hat s_z\sum_j\hat \sigma_z^{(j)}$ acts  in the $|\! \uparrow\rangle$ or $s_z=+1$ sector 
as a magnetic field $h=+g$ on each of the apparatus spins
and as a field $-g$ in the  $|\! \downarrow\rangle$ or $s_z=-1$ sector, this induces a Larmor rotation around the $z$-axis,
and, consequently, dephasing. For the elements $ r_{\uparrow\downarrow}^\ast $ and $r_{ \downarrow\uparrow}$
of $\hat r$, the density matrix of the tested spin, each apparatus spin brings a phase factor $\cos 2gt/\hbar$.
For the combination of the $N$ spins this brings  $\cos^N (2gt/\hbar)$.
This factor decays quickly, as $\exp(-t^2/\tau^2)$ with $\tau=\hbar/g\sqrt{2N}$ for large $N$. 
Unlike what is often stated, this is a {\it dephasing} effect, not a decoherence effect.
There is an analogy with NMR or MRI physics and  with $T_2$ relaxation, both being dephasing effects.

Decoherence by coupling to the environment, on the other hand, is the cause that the important phase information 
in MRI  after putting the spins in the same direction by a strong magnetic pulse, 
gets washed out if one waits too long before giving a $\pi$ pulse to restore the spins to their $t=0$ directions and record the accumulative effect.

Correlations between the tested spin and $k=1,2,3\ldots$ spins of the magnet, 
$\langle (\hat s_x\pm i\hat s_y)\hat \sigma_z^{(1)} \cdots \hat \sigma_z^{(k)} \rangle$
start at zero because S and A are uncorrelated at $t=0$, but grow a bit and then decay to zero  \cite{allahverdyan2013understanding}.
This demonstrates that the truncation $ r_{\uparrow\downarrow},\, r_{ \downarrow\uparrow}\to 0$
occurs in a cascade with multiparticle correlations between S and M: the information in S about $\hat s_x$, $\hat s_y$  ``leaks away into M''.

\subsection{Truncation of the density matrix of the apparatus by decoherence}

The previous cosine has undesired recurrences at $t_n=n\pi \hbar/g$. 
They are suppressed when the bath B sets in at a time $t_B$ well before $t_1$;
this effect, referred to as {\it decoherence}, means that the information on S leaks away in the bath (environment).
Then the contribution of each of the cosines is suppressed, so that the elements $\hat \cD_{\uparrow\downarrow}$ 
 and  $\hat \cD_{ \downarrow\uparrow}$ (defined in terms of $\hat \cD$ as in (3)) themselves are suppressed, 
 and with them their traces over A that yield $r_{\uparrow\downarrow}$ and  $r_{ \downarrow\uparrow}$, respectively.
This is termed ``truncation of the (off-diagonal elements of the) density matrix'' \cite{allahverdyan2005dynamics}, 
an essential ingredient for the later collapse interpretation of the density matrix,
i.e., the description of selecting individual systems with identical measurement outcome.

\subsection{Registration of the measurement}

\hfill{{\it Wie schrijft, die blijft}\footnote{Who writes, stays}}

\hfill{Dutch saying}

\vspace{3mm}

In the Curie--Weiss  model, in the $s_z=\pm 1$ sector,  the coupling acts as an external magnetic field of strength $g$ in the 
$\pm z$ direction on each of the apparatus spins.
The coupling strength $g$ has to be large enough to overcome the free energy barrier between the initial paramagnetic state and 
one of the thermodynamically stable states, so that a final magnetized state is reached.
If $g$ is too small, there is still a truncation but no registration, since after decoupling the system from the apparatus, 
the magnet will go back to its paramagnetic state.
These options are schematized  in figure \ref{figF}.

 \begin{figure}\label{figLoic1}
\centerline{ \includegraphics[width=8cm]{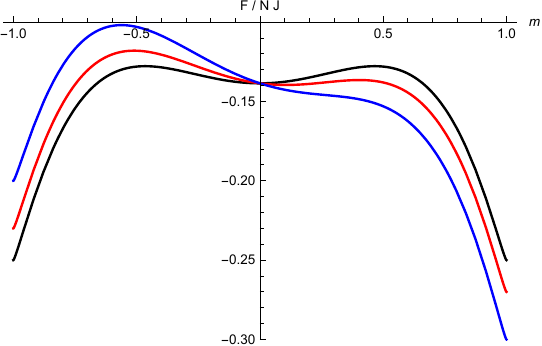}}
\caption{The free energy of the magnet in the sector $s_z=+1$ at $T=0.2J$ as function of normalized magnetization $m$.
For $t<0$ the coupling $g$ of S to A vanishes (black, symmetric curve). The magnet lies in its metastable paramagnetic state near $m=0$, while there are
stable states at $\pm m_F\approx\pm1$.
The barrier towards one of  the stable states (in this case, the one at $+m_F$)    is suppressed by a coupling $g>0.0357J$ 
so as to achieve registration of the measurement; depicted is the case $g=0.05J$ (blue curve).
For smaller $g$, like for $g=0.02J$ (red curve), no registration will be achieved, only truncation.\label{figF}
}
\end{figure}

The quantum dynamics can be expressed in $P(m,t)$, the distribution of the magnetization per magnet spin in the $z$-direction at time $t$. At $t=0$ 
this is a narrow peak around $m=0$ reflecting the initial paramagnetic state. In the course of time, the peak moves, widens and finally narrows
around the magnetized state $+ m_F$ in $r_{\uparrow\uparrow}$ and  $-m_F$ in $r_{\downarrow\downarrow}$, respectively.


\subsection{Post-measurement state}

\hfill{{\it After all was said and done,}

\hfill{\it  more was said than done}}

\hfill{Aesop}

\vspace{3mm}

At the end of the measurement, and putting $g\to0$ to decouple S from A, we end up with 
\BEQ
\hat{\cal  D}(t_{\rm f})= p_\uparrow |\uparrow\rangle\langle\uparrow| \otimes \cR _\Uparrow
+p_\downarrow |\downarrow\rangle\langle\downarrow|\otimes \cR _\Downarrow,
\qquad
p_\uparrow =r_{\uparrow\uparrow}(0),\quad
p_\downarrow =r_{\downarrow\downarrow}(0).
\EEQ
As discussed, the ``cat terms, viz, the $\uparrow\downarrow$ and $\downarrow\uparrow$ elements, have become irrelevant.
The shape is stable. Any composition of $\hat{\cal  D}(t_{\rm f})$ in two terms with the sum equal to  (4)
can be shown, in a microcanonical approach  but conserving the value of $m$,  to relax, for each of the terms, to a shape like (4), and hence for their combination.
This is due to small additional terms in the Curie-Weiss Hamiltonian, which cause a weak relaxational effect relevant only at this final stage.
The mechanism has been termed  subensemble relaxation in \cite{allahverdyan2013understanding}
 and polymicrocanonical relaxation in \cite{allahverdyan2017sub}. 
 
 \subsection{A run of experiments needs repeated resetting of the apparatus}

\hfill{\it Keep on running,}

\hfill{\it keep on hiding}

\hfill{Written by Jackie Edwards}

\hfill{Performed by the Spencer Davis Group}

\vspace{3mm}

 In an experiment the macroscopic pointer goes from its metastable ``ready''  state to one of its stable states.
 The difference in free energy is dumped in the bath and lost. To reset the apparatus for a next measurement, it has to be  brought back
 to its initial state, so one has to do work on it, which costs energy.
 This is the dominant reason why experiments are costly (and energy bills are a worry).
 
 In theoretical considerations of quantum foundations starting from the postulates, this physical aspect is not accounted for,
 showing how remote they are from what goes on in reality.
 
 \subsection{On the frequency interpretation of quantum mechanics}

Already in classical physics, the notion of probability has many interpretations, as discussed, for instance, in
one of Andrei's books \cite{khrennikov2009interpretations}. This dichotomy will not be solved -- but rather be deepened --
when going to the quantum level.
 
 In their ``Opusculo''  \cite{allahverdyan2017sub}, ABN consider the split of a general final density matrix into physical 
 sub-ensembles with post-measurement density matrices $\hat \cD_i$, viz.  $\hat\cD(t_{\it f})=\sum_i q_i\hat\cD_i$; 
  in the example (4) the sum over $i$ pertains to its two possible measurement outcomes.
   When joining two sub-ensembles with density matrices $\hat \cD^{(1,2)}$ and $N_{1,2}$ members, respectively,
 the new density matrix is
 \BEQ \hat{\cal D}=\lambda\hat{\cal D}^{(1)}+(1-\lambda)\hat{\cal D}^{(2)},
 \qquad \lambda=\frac{N_1}{N_1+N_2} .
 \EEQ
The weights $q_i$ of the respective splits obey  the rule
 \BEQ
 q_i=\lambda q_i^{(1)}+(1-\lambda)q_i^{(2)}.
 \EEQ
 This is the combination rule for frequencies in probability theory, so quantum theory directly embodies the frequency interpretation,
 the one that experimenters most often employ.
 The Bayesian interpretation of QM (QBism) adds priors of expectations to the quantum formalism \cite{fuchs2014introduction};
  it is then capable to give (additional) uncertainties in otherwise sharply defined theoretical values, such as employed at the LHC.

\subsection{Useless measurements are often still measurements}

Asher Peres says: unperformed measurements have no outcome. They are not only useless, they are meaningless.
Their repeated use in foundational considerations is a scandal, based on the misinterpretation of the postulates,
something that can not occur when modelling the physical apparatus and solving its dynamics.

But also performed measurements can be ``useless''. Imagine a Curie-Weiss apparatus -- to be discussed in detail in section 
\ref{Qmess-sec} below -- which is turned off after the truncation process but well before the registration of the measurement 
has been been achieved. The density matrix will have been truncated, the information coded in the off-diagonal elements 
will have been scrambled on its very short timescale.
But  the magnet will return to its paramagnetic initial state, and no registration will have taken place.
For practical purposes, this is a useless experiment. But (in the ensemble of measurements of this type,) 
forces have acted on the (the ensemble of) system(s) and have altered the density matrix, being the best coding of
knowledge about the ensemble.

Another case of such a ``useless'' measurement is when the system-apparatus coupling ($g$ in the Curie-Weiss model below)
is not strong enough to overcome the free energy barrier between the paramagnetic state and the stable, magnetized states.
After decoupling the apparatus from the system, the former will again go back to its paramagnetic state, and no sensible
pointer reading can be achieved. Also this is a ``useless'' experiment, in the sense of useless for testing elements of
the system's density matrix. But it is not useless if one merely wants to truncate it.

The best case of really useless experiments involves solar neutrinos. Each second our bodies are bombarded by hundreds of 
billions of them. Their coupling to atoms is so weak, that they they do not even achieve the truncation mechanism and can be left out of
modelling in QM. 
It is allowed to invoke them in idealized theoretical considerations, but they are of no help since they alter nothing.
Thank heavens, if they would couple stronger and really be able to interact with the atoms in our bodies, 
they would not only destroy life, but even have prevented it.

 \subsection{Measuring a ``classical''  system}

In order to understand the classical limit of quantum systems, it is instructive to consider the case where the also tested system itself is large,
consisting of $n\gg1$ spins $\hat{\bf s}^{(i)}$ for $i=1,2,\cdots,n$, coupled to some low temperature thermal bath, so that we consider an ensemble of 
systems magnetized upwards or downwards. Taking now as observable $\hat S=(1/n)\sum_{i=1}^n\hat s_z^{(i)}$ in the interaction Hamiltonian
$-g\hat S\hat M$, the previous $n=1$ theory can be generalized directly. 

In a first measurement, a fast truncation occurs over a time $(\hbar n/g)\sqrt{2/N}$,  are after which registration produces an equivalent of (4)
on the time scale $\hbar J/\gamma$ provided $n\ll g\sqrt{N}/\gamma J$. The would-be recurrences at $t_k=2nk\hbar /g$ come too late
to influence the pointer readings.
The decoherence should set in between them. Upon repeating the measurement an arbitrary number of times  in each of the 
up-up and down-down sectors, the previous outcome gets reproduced. 
The standard repeatability has for measuring a macroscopic system the smell of a classical system property: the moon is observed 
to be there ``again'' if you look again. This supports the individuality postulate of macroscopic systems in section \ref{IndivMacro}.

\subsection{Simultaneously measuring non-commuting variables: a model setup}

\hfill{{\it Je kunt niet op twee bruiloften tegelijk dansen}\footnote{You can't dance at two weddings simultaneously}}

\hfill{Dutch saying}

\vspace{3mm}

It is often said that non-commuting (incompatible) variables cannot be measured simultaneously; however,
 this statement is {\it incorrect}. The proper statement is: {\it when measuring incompatible variables simultaneously,
the Copenhagen postulates do not apply.}

In practice, such measurements are not impossible though difficult, as they involve at least two apparatuses, which will influence each other.
An idealized case has been modeled where both $\hat s_z$ and $\hat s_x$ of S, a spin $\half$,  are measured by two Curie-Weiss magnets:
 one, to be called M$_z$, having stable magnetized states in the $\pm \, z$-direction, and the other, to be called M$_x$, 
having them in the $\pm \, x$-direction \cite{allahverdyan2010simultaneous,allahverdyan2013understanding}.
The apparatuses will influence each other via their coupling to the common tested spin. 
The distribution of outcomes depends on many if not all details of the apparatuses;
if  M$_x$ is coupled to S before (or after) the M$_z$-apparatus is coupled, it also depends on the time delay between the coupling
instances, relaxing to the standard results in the limit of large time delay.

The case of zero time-delay between the coupling of the apparatuses to the spin,
the exactly simultaneous measurement, has been worked out in detail \cite{perarnau2017simultaneous}.
The probabilities for the four possible outcomes: magnetization of M$_z$ up or down, and magnetization of M$_x$ left or right, 
are explicit, depend on many system parameters and have no bearing on any Born rule that one could imagine for this case.
Moreover, there is no clear analog of the truncation of the density  matrix, or ``collapse of the wavefunction''.
But the measurement is still informative so as to allow reconstruction of the elements $\langle \hat s_{x}\rangle$ and $\langle \hat s_{z}\rangle$
of the initial density matrix of S from the outcomes of repeated measurements in this setup \cite{perarnau2017simultaneous}.
In the standard setup, this would require 2 sets of repeated measurements, for
measuring along the $x$ and $z$ directions, to determine 
$\langle \hat s_{x}\rangle$ and $\langle \hat s_{z}\rangle$ separately.

With a third apparatus, such that $\hat s_x$, $\hat s_y$ and $\hat s_z$ are  all measured simultaneously, 
one should be able to determine all elements of the density matrix of S from one (repeated) measurement. 
This task has so far not been worked out.

\section{Realistic quantum measurement as sine-qua-non basis for interpretation}
\label{sinequanon}

\hfill{\it I will liken him unto a wise man,}

\hfill{\it which built his house upon a rock}

\hfill{Matthew VII:24}

\vspace{3mm}

The project of interpretation of QM of course goes all the way back to the introduction of this theory.
Classical papers on interpretation of QM and quantum measurement have been collected in \cite{wheeler2014quantum}.
Lacking a convincing solvable model, the opinions remained divergent.

Like the proof of the pudding is in the eating,  
the only point of contact between quantum theory and the reality in the laboratory lies in quantum measurements.
Hence interpretation should be based on analyzing measurement in a specified setup (context),
which can be an optical table with beam splitters, mirrors, phase delayers, photon counters, and so on.
Together they form the context for the measurement and interpretation needs only refer to a given context;
all further attempts can not offer a way to test them.
Grangier and Auff\`eves stress this by introducing the term of the ``modality'' for the quantum system in its specified context\cite{auffeves2016contexts}.
Other interpretations like 
Wigner's friend  \cite{wheeler2014quantum,laloe2019we} and relational QM \cite{rovelli1996relational} try to grasp the contextuality, 
bounded more by the imagination of the human mind than by the  stringent structure of the quantum formalism.

Following Bohr, we consider QM as a theory that makes predictions in a well defined context.
The role of context in ordinary life is well known. For example, a joke which does well in one context, can be sour in another. 
Even the same word in the same language  can have a different meaning in a different context.

With the solution of the Curie-Weiss model at hand, ABN generalize this for a class of similar models and interpret the results
in a minimalistic fashion.
The main points are: 

\begin{itemize}
\item The natural interpretation of QM is a specified version of the statistical interpretation. QM describes ensembles of
identically prepared systems or particles, tested in an ensemble of measurements in a specified context.

\item The density matrix is an abstract mathematical object, which codes our best knowledge about the ensemble.
It needs no interpretation (no ``wave function of the particle''). It leads to an epistemology without connection to an ontology,
it describes statistics but is silent about the individual events.

\item Results from measurements in different contexts should not be combined, 
for instance because apparatuses can not stand in each others way.

\item Individual events do exist, but quantum theory can only treat them statistically.

\item 
Truncation of the density matrix of the tested system
(disappearance of off-diagonal ``cat'' terms terms) occurs by dephasing followed by decoherence.
The latter also erases the related terms for the apparatus. 

\item The registration of the measurement is related to a dynamical phase transition 
from the initial ``ready'' metastable thermodynamic phase to one of the stable thermodynamic phases.
This transition amplifies the small quantum signal and causes finally a macroscopic effect in the apparatus, which can be read off 
or processed electronically.

\item
The system-apparatus coupling has to be large enough to overcome the free energy barrier between the initial metastable 
and final stable state. The excess free energy is dumped into the bath. Resetting the apparatus costs energy.

\item Ideal measurements constitute the mathematical limit of a series of more and more idealized physical measurements.
Interpretation of the quantum formalism follows from considering suitable ideal measurements.

\item
The Born rule follows from the quantum formalism itself, namely from a dynamical conservation law in ideal measurements.

\item
Born probabilities can be related to the macroscopic pointer readings, with which the tested microscopic system is 
correlated\footnote{This ``view from the top'' is opposite to the Copenhagen interpretation, which focusses on the tested system and treats 
the apparatus in an abstract way by a postulate.}.
Non-idealities lead to false readings.

\item
Noncommuting variables can be measured simultaneously. The outcome probabilities
are not described by the Born rule, but involve the rich physics of the apparatuses, which get coupled to each other 
via the commonly tested system.
\end{itemize}

\section{The elephant in the room: no theory for individual events}
\label{elephant}


\hfill{\it This emptiness deep inside}

\hfill{Expression}

\vspace{3mm}

There is no working model for the shape of an individual electron.
Quantum electrodynamics states it to be a point particle with a quantum cloud of virtual particles and antiparticles around it.
But point particles do not exist, they are  an abstraction for a description of some small entity, so we still dream of a 
better modelling.

In our view, particles do exist (virtual particles live shortly). 
Clear evidences are solid state materials and high energy cosmic rays.
QM provides the Born probabilities for ensembles of measurements on ensembles of particles.
But what about the individual measurements? What happens in them? It is our thesis that we have no theory for it, no ontology!
This is ``a problem as big as an elephant'' or, said otherwise, the proverbial elephant in the room.

\subsection{The EPR setup}

\hfill{\it Wanting to beat Einstein}

\hfill{\it equates to}

\hfill{{\it Being alive and kicking}

\hfill{Anonymous}

\vspace{3mm}

In an Einstein-Podolsky-Rosen (EPR) experiment for a pair of, say, electrons in an $S=0$ Bell state, measurement at A along 
an arbitrary unit vector ${\bf n}$,
will give an outcome $s_A=\pm 1$ (in units of $\half\hbar$). Measurement at the opposite station B along the same
axis ${\bf n}$ will result in $s_B=\mp1$, so that the sum leads to $s_A+s_B=0$. 
This is a manifestation of  the conservation law of total spin in this sector, ${\bf S}\cdot{\bf n}=0$. 
Whether the experiment at $A$ is performed
before or after the one at $B$ is set by the distance from the source to the respective detectors. They can 
be such that the second experiment is far enough that it gets performed before a light signal from the first could arrive.

In lack of a subquantum theory that would clarify the situation, it has been proposed that there has to be 
faster-than-light communication between the particles. This poses some questions.
How is that compatible with the relativity principle?
Would the information be sent when the first particle is measured, to the second, carried by it till that also gets measured?
Or sent to the second apparatus to be employed when its particle arrives? 
No detailed model is known, and no definite answer for this conundrum is agreed upon.

We are left with the question: How does Nature get it done? 
 A simple model is that the information for the measurement outcomes travels with the particles, so that they can instruct the detectors. 
 A role is then likely played by the clouds of virtual quantum  particles. 
 For this much thinking will be needed, but the goal is assured by the existence of the individual events.

\subsection{Wanted living or alive: Subquantum mechanics as a less statistical theory}

One may consider that the quantum uncertainty arrives from averaging over some very fast underlying dynamics,
with quantum theory representing the behavior at the longer timescales, the standard case of time scale separation in glasses\cite{leuzzi2007thermodynamics}.
A consistent research line based on cellular automata is developed by Gerard 't Hooft \cite{t2016cellular}.
For the time being one can be amazed that quantum mechanics captures the essentials of this subquantum physics,
partly as conservation laws, partly in a statistical way. 

To me Einstein's dream of a reality is as vivid as it ever was.
The last but not least  of the scandals of quantum mechanics is that, after one century, there is still this most fundamental question to answer!
This is unbearable, so Andrei: solve it soon. Or else: please organize many further V\"axj\"o meetings, so that we can attack it as a community!


\begin{thebibliography}{10}
\providecommand{\url}[1]{\texttt{#1}}
\providecommand{\urlprefix}{URL }
\expandafter\ifx\csname urlstyle\endcsname\relax
  \providecommand{\doi}[1]{doi:\discretionary{}{}{}#1}\else
  \providecommand{\doi}{doi:\discretionary{}{}{}\begingroup
  \urlstyle{rm}\Url}\fi
\providecommand{\eprint}[2][]{\url{#2}}

\bibitem{allahverdyan2005brownian}
A.~E. Allahverdyan, A.~Khrennikov and T.~M. Nieuwenhuizen,
\newblock \emph{Brownian entanglement},
\newblock Physical Review A \textbf{72}(3), 032102 (2005).

\bibitem{van2008scandal}
N.~Van~Kampen,
\newblock \emph{The scandal of quantum mechanics},
\newblock American Journal of Physics \textbf{76}(11), 989 (2008).

\bibitem{nieuwenhuizen1987objections}
T.~M. Nieuwenhuizen, D.~Frenkel and N.~Van~Kampen,
\newblock \emph{Objections to Handel’s quantum theory of 1/f noise},
\newblock Physical Review A \textbf{35}(6), 2750 (1987).

\bibitem{van1988ten}
N.~Van~Kampen,
\newblock \emph{Ten theorems about quantum mechanical measurements},
\newblock Physica A: Statistical Mechanics and its Applications
  \textbf{153}(1), 97 (1988).

\bibitem{peres1978unperformed}
A.~Peres,
\newblock \emph{Unperformed experiments have no results},
\newblock American Journal of Physics \textbf{46}(7), 745 (1978).

\bibitem{plotnitsky2011reasonable}
A.~Plotnitsky,
\newblock \emph{On the reasonable and unreasonable effectiveness of mathematics
  in classical and quantum physics},
\newblock Foundations of Physics \textbf{41}(3), 466 (2011).

\bibitem{khrennikov2014ubiquitous}
A.~Y. Khrennikov,
\newblock \emph{Ubiquitous quantum structure},
\newblock Springer (2014).

\bibitem{allahverdyan2013understanding}
A.~E. Allahverdyan, R.~Balian and T.~M. Nieuwenhuizen,
\newblock \emph{Understanding quantum measurement from the solution of
  dynamical models},
\newblock Physics Reports \textbf{525}(1), 1 (2013).

\bibitem{darrigol2015some}
O.~Darrigol,
\newblock \emph{Why some physical theories should never die},
\newblock {\'E}vora Studies in the Philosophy and History of Science. In
  memoriam Herm{\'\i}nio Martins pp. 319--368 (2015).

\bibitem{auffeves2016contexts}
A.~Auff{\`e}ves and P.~Grangier,
\newblock \emph{Contexts, systems and modalities: a new ontology for quantum mechanics},
\newblock Foundations of Physics \textbf{46}(2), 121 (2016).

\bibitem{khrennikov2014beyond}
A.~Khrennikov,
\newblock \emph{Beyond quantum},
\newblock CRC Press (2014).

\bibitem{plotnitsky2015reality}
A.~Plotnitsky and A.~Khrennikov,
\newblock \emph{Reality without realism: on the ontological and epistemological
  architecture of quantum mechanics},
\newblock Foundations of Physics \textbf{45}(10), 1269 (2015).

\bibitem{nieuwenhuizen1983exact}
T.~M. Nieuwenhuizen,
\newblock \emph{Exact electronic spectra and inverse localization lengths in
  one-dimensional random systems: I. random alloy, liquid metal and liquid
  alloy},
\newblock Physica A: Statistical Mechanics and its Applications
  \textbf{120}(3), 468 (1983).

\bibitem{de1985distribution}
C.~de~Calan, J.-M. Luck, T.~M. Nieuwenhuizen and D.~Petritis,
\newblock \emph{On the distribution of a random variable occurring in 1d
  disordered systems},
\newblock Journal of Physics A: Mathematical and General \textbf{18}(3), 501
  (1985).

\bibitem{nieuwenhuizen1989trapping}
T.~M. Nieuwenhuizen,
\newblock \emph{Trapping and Lifshitz tails in random media, self-attracting
  polymers, and the number of distinct sites visited: a renormalized instanton
  approach in three dimensions},
\newblock Physical Review Letters \textbf{62}(4), 357 (1989).

\bibitem{vRossum1993band}
M.~van Rossum, T.~M. Nieuwenhuizen, E.~Hofstetter and M.~Schreiber,
\newblock \emph{Band tails in a disordered system},
\newblock In C.~Soukoulis, ed., \emph{Photonic Band Gaps and Localization}, pp.
  509--513. Plenum Press, New York (1993).

\bibitem{van1994density}
M.~Van~Rossum, T.~M. Nieuwenhuizen, E.~Hofstetter and M.~Schreiber,
\newblock \emph{Density of states of disordered systems},
\newblock Physical Review B \textbf{49}(19), 13377 (1994).

\bibitem{den1993location}
P.~Den~Outer, T.~M. Nieuwenhuizen and A.~Lagendijk,
\newblock \emph{Location of objects in multiple-scattering media},
\newblock JOSA A \textbf{10}(6), 1209 (1993).

\bibitem{de1994probability}
J.~F. de~Boer, M.~Van~Rossum, M.~P. van Albada, T.~M. Nieuwenhuizen and
  A.~Lagendijk,
\newblock \emph{Probability distribution of multiple scattered light measured
  in total transmission},
\newblock Physical Review Letters \textbf{73}(19), 2567 (1994).

\bibitem{nieuwenhuizen1993skin}
T.~M. Nieuwenhuizen and J.~Luck,
\newblock \emph{Skin layer of diffusive media},
\newblock Physical Review E \textbf{48}(1), 569 (1993).

\bibitem{nieuwenhuizen1995intensity}
T.~M. Nieuwenhuizen and M.~Van~Rossum,
\newblock \emph{Intensity distributions of waves transmitted through a multiple
  scattering medium},
\newblock Physical Review Letters \textbf{74}(14), 2674 (1995).

\bibitem{van1999multiple}
M.~van Rossum and T.~M. Nieuwenhuizen,
\newblock \emph{Multiple scattering of classical waves: microscopy, mesoscopy,
  and diffusion},
\newblock Reviews of Modern Physics \textbf{71}(1), 313 (1999).

\bibitem{nieuwenhuizen1995exactly}
T.~M. Nieuwenhuizen,
\newblock \emph{Exactly solvable model of a quantum spin glass},
\newblock Physical Review Letters \textbf{74}(21), 4289 (1995).

\bibitem{nieuwenhuizen1997ehrenfest}
T.~M. Nieuwenhuizen,
\newblock \emph{Ehrenfest relations at the glass transition: solution to an old
  paradox},
\newblock Physical Review Letters \textbf{79}(7), 1317 (1997).

\bibitem{nieuwenhuizen1998thermodynamics}
T.~M. Nieuwenhuizen,
\newblock \emph{Thermodynamics of the glassy state: effective temperature as an
  additional system parameter},
\newblock Physical Review Letters \textbf{80}(25), 5580 (1998).

\bibitem{leuzzi2007thermodynamics}
L.~Leuzzi and T.~M. Nieuwenhuizen,
\newblock \emph{Thermodynamics of the glassy state},
\newblock CRC Press (2007).

\bibitem{allahverdyan2004maximal}
A.~E. Allahverdyan, R.~Balian and T.~M. Nieuwenhuizen,
\newblock \emph{Maximal work extraction from finite quantum systems},
\newblock EPL (Europhysics Letters) \textbf{67}(4), 565 (2004).

\bibitem{jaynes2003probability}
E.~T. Jaynes,
\newblock \emph{Probability theory: The logic of science},
\newblock Cambridge University Press (2003).

\bibitem{balian2004entropy}
R.~Balian,
\newblock \emph{Entropy, a protean concept},
\newblock Progress in Mathematical Physics \textbf{38}, 119 (2004).

\bibitem{balian2005information}
R.~Balian,
\newblock \emph{Information in statistical physics},
\newblock Studies in History and Philosophy of Science Part B: Studies in
  History and Philosophy of Modern Physics \textbf{36}(2), 323 (2005).

\bibitem{brillouin2014scientific}
L.~Brillouin,
\newblock \emph{Scientific uncertainty and information},
\newblock Academic Press (2014).

\bibitem{gibbs1906scientific}
J.~W. Gibbs,
\newblock \emph{Scientific Papers of J. Willard Gibbs, in Two Volumes}, vol.~1,
\newblock Longmans, Green (1906).

\bibitem{allahverdyan2006explanation}
A.~E. Allahverdyan and T.~M. Nieuwenhuizen,
\newblock \emph{Explanation of the Gibbs paradox within the framework of
  quantum thermodynamics},
\newblock Physical Review E \textbf{73}(6), 066119 (2006).

\bibitem{allahverdyan2021dynamical}
A.~E. Allahverdyan, K.~V. Hovhannisyan and D.~Petrosyan,
\newblock \emph{Dynamical symmetrization of the state of identical particles},
\newblock Proceedings of the Royal Society A \textbf{477}(2249), 20200911
  (2021).

\bibitem{allahverdyan2017sub}
A.~E. Allahverdyan, R.~Balian and T.~M. Nieuwenhuizen,
\newblock \emph{A sub-ensemble theory of ideal quantum measurement processes},
\newblock Annals of Physics \textbf{376}, 324 (2017).

\bibitem{wikiWheeler}
wikipedia,
\newblock \emph{One-electron universe},
\newblock Recovered on Dec 1  (2021),
\newblock \doi{https://en.wikipedia.org/wiki/One-electron_universe}.

\bibitem{de2014quantum}
H.~De~Raedt, M.~I. Katsnelson and K.~Michielsen,
\newblock \emph{Quantum theory as the most robust description of reproducible
  experiments},
\newblock Annals of Physics \textbf{347}, 45 (2014).

\bibitem{de2015quantum}
H.~De~Raedt, M.~I. Katsnelson, H.~C. Donker and K.~Michielsen,
\newblock \emph{Quantum theory as a description of robust experiments:
  Derivation of the pauli equation},
\newblock Annals of Physics \textbf{359}, 166 (2015).

\bibitem{auffeves2020deriving}
A.~Auff{\`e}ves and P.~Grangier,
\newblock \emph{Deriving Born’s rule from an inference to the best
  explanation},
\newblock Foundations of Physics \textbf{50}(12), 1781 (2020).

\bibitem{khrennikov2003contextual}
A.~Khrennikov,
\newblock \emph{Contextual viewpoint to quantum stochastics},
\newblock Journal of Mathematical Physics \textbf{44}(6), 2471 (2003).

\bibitem{wheeler2014quantum}
J.~A. Wheeler and W.~H. Zurek,
\newblock \emph{Quantum theory and measurement}, vol.~53,
\newblock Princeton University Press (2014).

\bibitem{bohm1952suggested}
D.~Bohm,
\newblock \emph{A suggested interpretation of the quantum theory in terms of"
  hidden" variables. i},
\newblock Physical Review \textbf{85}(2), 166 (1952).

\bibitem{holland1995quantum}
P.~R. Holland,
\newblock \emph{The quantum theory of motion: an account of the de Broglie-Bohm
  causal interpretation of quantum mechanics},
\newblock Cambridge university press (1995).

\bibitem{zeh1970interpretation}
H.~D. Zeh,
\newblock \emph{On the interpretation of measurement in quantum theory},
\newblock Foundations of Physics \textbf{1}(1), 69 (1970).

\bibitem{joos2013decoherence}
E.~Joos, H.~D. Zeh, C.~Kiefer, D.~J. Giulini, J.~Kupsch and I.-O. Stamatescu,
\newblock \emph{Decoherence and the appearance of a classical world in quantum
  theory},
\newblock Springer Science \& Business Media (2013).

\bibitem{zurek2003decoherence}
W.~H. Zurek,
\newblock \emph{Decoherence, einselection, and the quantum origins of the
  classical},
\newblock Reviews of Modern Physics \textbf{75}(3), 715 (2003).

\bibitem{allahverdyan2003curie}
A.~E. Allahverdyan, R.~Balian and T.~M. Nieuwenhuizen,
\newblock \emph{Curie-weiss model of the quantum measurement process},
\newblock EPL (Europhysics Letters) \textbf{61}(4), 452 (2003).

\bibitem{allahverdyan2003thequantum}
A.~E. Allahverdyan, R.~Balian and T.~M. Nieuwenhuizen,
\newblock \emph{The quantum measurement process: an exactly solvable model},
  vol. An. LVIII,
\newblock \doi{https://arxiv.org/abs/cond-mat/0309188} (2003).

\bibitem{allahverdyan2004dynamics}
A.~E. Allahverdyan, R.~Balian and T.~M. Nieuwenhuizen,
\newblock \emph{Dynamics of quantum measurements}  (2004),
\newblock \doi{https://arxiv.org/pdf/quant-ph/0412045.pdf}.

\bibitem{allahverdyan2005quantum}
A.~E. Allahverdyan, R.~Balian and T.~M. Nieuwenhuizen,
\newblock \emph{The quantum measurement process in an exactly solvable model},
\newblock In \emph{AIP Conference Proceedings}, vol. 750, pp. 26--34. American
  Institute of Physics (2005).

\bibitem{allahverdyan2006phase}
A.~E. Allahverdyan, R.~Balian and T.~M. Nieuwenhuizen,
\newblock \emph{Phase transitions and quantum measurements},
\newblock In \emph{AIP Conference Proceedings}, vol. 810, pp. 47--58. American
  Institute of Physics (2006).

\bibitem{nieuwenhuizen2014lectures}
T.~M. Nieuwenhuizen, M.~Perarnau-Llobet and R.~Balian,
\newblock \emph{Lectures on dynamical models for quantum measurements},
\newblock International Journal of Modern Physics B \textbf{28}(21), 1430014
  (2014).

\bibitem{khrennikov2008born}
A.~Khrennikov,
\newblock \emph{Born's rule from classical random fields},
\newblock Physics Letters A \textbf{372}(44), 6588 (2008).

\bibitem{allahverdyan2005dynamics}
A.~E. Allahverdyan, R.~Balian and T.~M. Nieuwenhuizen,
\newblock \emph{Dynamics of a quantum measurement},
\newblock Physica E: Low-dimensional Systems and Nanostructures
  \textbf{29}(1-2), 261 (2005).

\bibitem{khrennikov2009interpretations}
A.~Khrennikov,
\newblock \emph{Interpretations of probability},
\newblock de Gruyter (2009).

\bibitem{fuchs2014introduction}
C.~A. Fuchs, N.~D. Mermin and R.~Schack,
\newblock \emph{An introduction to qbism with an application to the locality of
  quantum mechanics},
\newblock American Journal of Physics \textbf{82}(8), 749 (2014).

\bibitem{allahverdyan2010simultaneous}
A.~E. Allahverdyan, R.~Balian and T.~M. Nieuwenhuizen,
\newblock \emph{Simultaneous measurement of non-commuting observables},
\newblock Physica E: Low-dimensional Systems and Nanostructures \textbf{42}(3),
  339 (2010).

\bibitem{perarnau2017simultaneous}
M.~Perarnau-Llobet and T.~M. Nieuwenhuizen,
\newblock \emph{Simultaneous measurement of two noncommuting quantum variables:
  Solution of a dynamical model},
\newblock Physical Review A \textbf{95}(5), 052129 (2017).

\bibitem{laloe2019we}
F.~Lalo{\"e},
\newblock \emph{Do we really understand quantum mechanics?},
\newblock Cambridge University Press (2019).

\bibitem{rovelli1996relational}
C.~Rovelli,
\newblock \emph{Relational quantum mechanics},
\newblock International Journal of Theoretical Physics \textbf{35}(8), 1637
  (1996).

\bibitem{t2016cellular}
G.~'t~Hooft,
\newblock \emph{The cellular automaton interpretation of quantum mechanics},
\newblock Springer Nature (2016).

\end{thebibliography}
\end{document}